\documentclass[dvipsnames]{svmult}

\usepackage{type1cm}        
\usepackage{makeidx}         
\usepackage{graphicx}        
\usepackage{multicol}        
\usepackage[bottom]{footmisc}
\usepackage{amsfonts}

\newcommand{\lnk}[1]{\texttt{\textcolor{Blue}{\url{#1}}}}
\newcommand{\ttl}[1]{\texttt{\textcolor{MidnightBlue}{#1}}}

\newcommand{\cf}{\textit{cf.}}
\newcommand{\eg}{\textit{e.g.}}
\newcommand{\etal}{\textit{et al.}}
\newcommand{\etc}{\textit{etc.}}
\newcommand{\ie}{\textit{i.e.}}
\newcommand{\nb}{\textit{n.b.}}

\makeindex             

\begin{document}

\title*{Reliable and interoperable computational molecular engineering: 2.~Semantic interoperability based on the European Materials and Modelling Ontology}
\titlerunning{Semantic interoperability based on the EMMO}
\author{Martin Thomas Horsch, Silvia Chiacchiera, Youness Bami, Georg J.\ Schmitz, Gabriele Mogni, Gerhard Goldbeck, and Emanuele Ghedini}
\authorrunning{Horsch, Chiacchiera, Bami, Schmitz, Mogni, Goldbeck, Ghedini}
\institute{Martin Thomas Horsch and Silvia Chiacchiera \at UK Research and Innovation, STFC Daresbury Laboratory, Keckwick Ln, Daresbury, Cheshire WA4~4AD, UK, \email{\{martin.horsch, silvia.chiacchiera\}@stfc.ac.uk}
\and Youness Bami and Georg J.\ Schmitz \at ACCESS e.V., Intzestr.\ 5, 52072 Aachen, Germany, \email{\{y.bami, g.j.schmitz\}@access.rwth-aachen.de}
\and Gabriele Mogni and Gerhard Goldbeck \at Goldbeck Consulting Ltd, St John's Innovation Centre, Cowley Rd, Cambridge CB4~0WS, UK, \email{\{gabriele, gerhard\}@goldbeck-consulting.com}
\and Emanuele Ghedini \at Universit\`a di Bologna, Department of Industrial Engineering, Via Saragozza 8, 40123, Bologna, Italy, \email{emanuele.ghedini@unibo.it}}

\maketitle

\abstract*{The European Materials and Modelling Ontology (EMMO) is a top-level ontology that was designed by the European Materials Modelling Council (EMMC) to facilitate semantic interoperability between platforms, models, and tools in computational molecular engineering, integrated computational materials engineering, and related applications of materials modelling and characterization. Additionally, domain ontologies exist based on data technology developments from specific platforms. The present work discusses the ongoing work on establishing a European Virtual Marketplace Framework, into which diverse platforms can be integrated. It addresses common challenges that arise when marketplace-level domain ontologies are combined with a top-level ontology like the EMMO by ontology alignment.}

\abstract{The European Materials and Modelling Ontology (EMMO) is a top-level ontology that was designed by the European Materials Modelling Council (EMMC) to facilitate semantic interoperability between platforms, models, and tools in computational molecular engineering, integrated computational materials engineering, and related applications of materials modelling and characterization. Additionally, domain ontologies exist based on data technology developments from specific platforms. The present work discusses the ongoing work on establishing a European Virtual Marketplace Framework, into which diverse platforms can be integrated. It addresses common challenges that arise when marketplace-level domain ontologies are combined with a top-level ontology like the EMMO by ontology alignment.}

\section{Introduction}
\label{sec:intro}

Semantic interoperability refers to an agreement between multiple
software or data infrastructures on
the terms by which a scenario, \ie, the circumstances
relevant to a particular application, can be described. In the
presence of such an agreement, different platforms can annotate
data with metadata of the same type and associated to each other
by the same relations, so that their meaning is clarified and an
exchange of information can rely on standardized \textit{semantics}.
This is independent of the precise technical mode in which the
exchange is implemented, which additionally requires an agreement
at the syntactic level, \eg, on the file formats or application programming
interfaces that are employed for this purpose. There are
common approaches by which semantic content can be serialized (\ie, denoted) straightforwardly,
\eg, using the JSON, YAML, TTL, and RDF/XML formats; obversely, semantic information
can be included in file formats that permit the annotation of their content,
such as CML~\cite{WMCTBTWDAP06, XRGP19} and HDF5~\cite{DCH14, Schmitz16}.
Accordingly, if there is a common standard for the semantics, solutions
for achieving syntactic interoperability are immediately available as well.
Semantic standards are documented by \textit{semantic assets},
which include, in increasing order of expressivity~\cite{EMMC19b},
\begin{enumerate}
   \item{} simple lists, where labels or names for concepts are collected without any further elaboration,
   \item{} informal hierarchies, which include further information on the concepts,
   \item{} thesauri, where concept definitions are presented systematically in text form, possibly including certain relations, \eg, between synonymous or antonymous concepts,
   \item{} taxonomies, \ie, class hierarchies,
   \item{} conceptual models, \eg, XSD or RDFS schemas, which extend the class hierarchy by a specification of possible contents or relations,
   \item{} ontologies, \ie, class hierarchies + relation definitions + axioms.
\end{enumerate}
This is known as the hierarchy of semantic assets; levels 1 to 3
are targeted at human users, whereas at levels 4 to 6, a machine-processable
formalization is usually given.
This work deals with \textit{ontologies} which, beside defining
classes and their properties in a way largely analogous to object-oriented
programming, also include inference rules (or axioms) that can be
used for automated reasoning. The OWL language, for which a
variety of formats exist, is the established standard for
formulating these definitions and rules; thereby, references to
concepts from other ontologies or metadata schemas can be included,
by which a \textit{semantic web} is created in a similar way as
hyperlinks are fundamental to the world wide web~\cite{AH11}.

The technical use of ontologies typically requires an agreement on
data and metadata related to one or multiple specific fields, or
domains of knowledge, for which domain ontologies need to be
developed in an effort requiring competencies from data
management as well as concrete scientific or technical expertise.
In materials informatics, a variety of
ontology-based approaches have been employed
successfully to integrate data from diverse sources~\cite{MMCMGP17, ZQ17}.
Moreover, domain-specific automated reasoning mechanisms can enhance
the functionality of cyber-physical and industry 4.0 infrastructures,
supplementing the evaluation of discrete process models, such as Petri nets~\cite{WWJ19}.

Since ontologies can point to other ontologies, the
semantic web exhibits the tendency for the semantic assets to aggregate into
large connected networks with a complex structure.
If algorithmic logical reasoning is
applied to these structures, any inconsistencies become fatal, since a
single contradiction at any point is sufficient to destroy the
logical coherence of the entire content as a whole.
On the other hand, it is practically
impossible to avoid mutual inconsistencies between any out of the
multitude of domain ontologies that are developed by different groups,
coming from different fields, who may all
have their own approach to the problem.
This raises the challenge
of establishing semantic standards that apply at highest possible level of
abstraction, under which all conceivable domain ontologies can be subsumed;
these components, the design of which requires a solid philosophical
underpinning, are known as \textit{top-level ontologies} or upper
ontologies~\cite{ASS15, BM10, MLR10}.
The present work addresses the problem of making a top-level
ontology viable technically by connecting it to domain-specific
ontologies and scenarios. The considered top-level ontology is the
European Materials and Modelling Ontology (EMMO)~\cite{EMMC19}, which is developed
by the European Materials Modelling Council (EMMC) on the basis of
previous efforts, \cf~Schmitz~\etal~\cite{SBAELABBV16}. The relevant
domains of knowledge are related to applications of materials modelling,
including computational molecular engineering (CME), which has a focus
on fluids and interfacial phenomena~\cite{HNVH13, HNBCSCELNSSTVC20},
and integrated computational materials engineering (ICME),
which has a focus on solids~\cite{Schmitz16, SBAELABBV16}.

Knowledge bases that employ semantic technology can be split into two components, one of which is terminological and deals with universals (here, \textit{classes}) and possible worlds, \ie, what relations there can theoretically be between objects (here, \textit{individuals}); the second one is assertional and deals with a factual scenario, \ie, a description of the real world as it is, or with a concrete possible scenario that is not necessarily factual. Accordingly, the assertional component contains individuals and defines actual (rather than merely possible or necessary) relations between them. In description logic, these two components are referred to as the TBox and the ABox, and in model theory, the content of the ABox is referred to as a model \cite{BHLS17}; here we refrain from using this term to avoid confusion with the concept of a model in materials modelling in general and the EMMO in particular. Instead, the term \textit{ontology} will be used specifically for the terminological part (including axioms and rules), and \textit{scenario} for a possible, concrete realization of the assertional part; this understanding of the term is in line with the philosophical paradigm of nominalism, \ie, non-existence of universals \cite{BP15, Lewis83, PM12}, and its implications on ontology design as implemented by the EMMO, which generally avoids the explicit definition of individuals. This yields a strict separation between terminology (ontology), which applies to classes, and assertions that apply to individuals. Accordingly, where an ontology definition file (\eg, in TTL format) includes individuals, as it is the case for the present domain ontologies, \cf~Section~\ref{sec:marketplace-level}, the latter are considered to be part of the scenario, not the ontology.

The remainder of the present work is structured as follows: Section~\ref{sec:marketplace-level} introduces ontologies used by a system of interoperable infrastructures for CME/ICME services: The European Virtual Marketplace Framework. Section \ref{sec:combining} discusses the problem of matching top-level and domain ontologies with each other, from a general point of view, and Section \ref{sec:example} applies this to the challenge of describing concrete features of services provided at the European Virtual Marketplace Framework in agreement with the joint top-level interoperability standard which is given by the EMMO; to illustrate this, a scenario from molecular modelling is considered in detail. Conclusions are formulated in Section~\ref{sec:conclusion}. Work in progress on representing EMMO scenarios as graphs and using these graphs to automatically create Python objects is documented in the Appendix.

\section{Semantic assets for services and marketplace platforms}
\label{sec:marketplace-level}

\subsection{European Virtual Marketplace Framework}
\label{subsec:evmpo}

The European Virtual Marketplace Framework, established by the joint work of the MarketPlace and Virtual Materials Marketplace (VIMMP) consortia in coordination with the EMMC, is open to participation by any interested provider, translator (\ie, facilitator), or end user of CME/ICME services. It is entirely based on transparent and openly accessible specifications, in particular, on a coherent system of ontologies with the EMMO at the top level. By creating an open framework on the basis of community-governed interoperability standards, a variety of projects, many of which (including EMMC-CSA, MarketPlace, VIMMP, and SimDOME) are funded from the LEIT-NMBP line of the European Union's Horizon 2020 research and innovation programme, contribute to a system of platforms and infrastructures that will support the uptake of CME/ICME solutions by industrial research and development practice.

The European Virtual Marketplace Ontology (EVMPO) was agreed as a common point of departure for the standardization of service-oriented semantics, relevant to marketplace platforms, by the projects involved in establishing the European Virtual Marketplace Framework. By defining eleven funda­mental paradigmatic categories, which correspond to irreducible terms that are constitutive to the paradigm underlying \textit{materials model\-ling marketplaces}, the EVMPO provides a structure for further, more specific marketplace-level ontologies. These fundamental paradigmatic categories are specified in Section~\ref{subsec:paradigmatic}.

\begin{figure}[p]
\centering
\includegraphics[width=9cm]{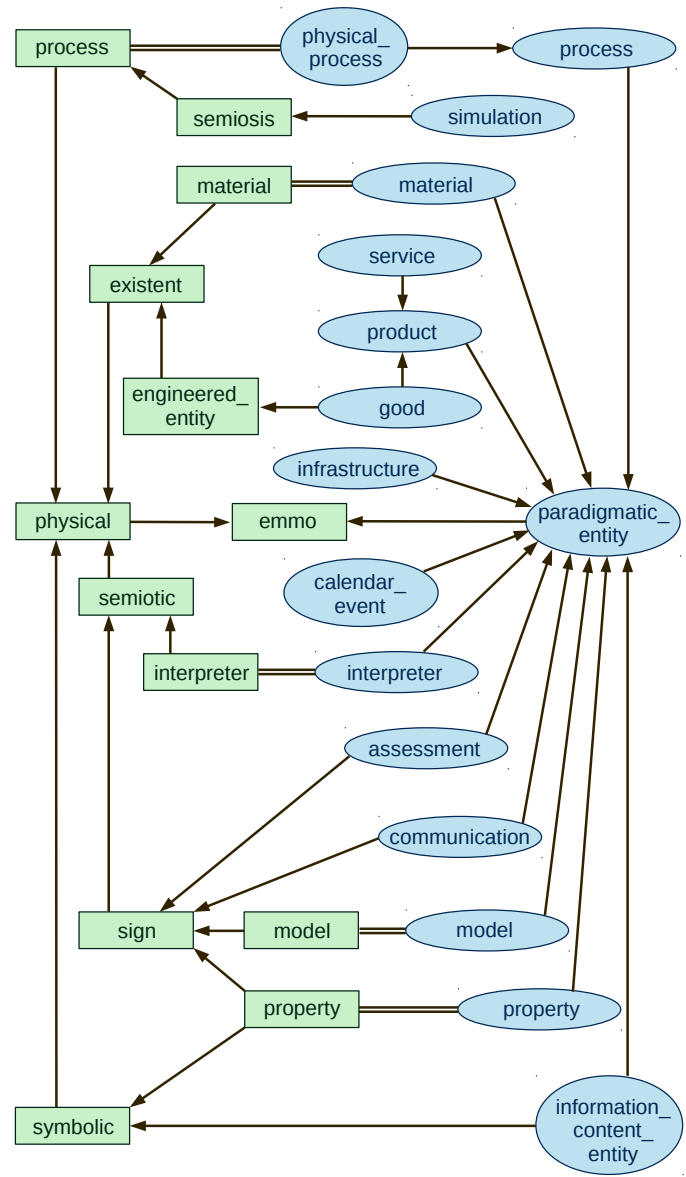}
\caption{Fundamental paradigmatic categories and other classes from the EVMPO (ellipses) together with related classes from the EMMO (rectangles); arrows between classes represent the transitive reduction of the \ttl{rdfs:subClassOf} relation, and double lines between classes represent the \ttl{owl:equivalentClass} relation. Note: The EVMPO and the EMMO are work in progress; this diagram refers to EMMO pre-release version 0.9.10 \cite{EMMC19}.}
\label{fig:emmo-evmpo}
\end{figure}

Terms which are not closely related to the marketplace paradigm itself, but occur in the related semantic web, are defined to be \textit{non-paradigmatic}. For this purpose, the EVMPO includes \ttl{evmpo:annotation} as a fundamental non-paradigmatic category. Below the fundamental level, the EVMPO also includes non-fundamental entites as subclasses, \eg, \ttl{evmpo:simulation} as a subclass of \ttl{evmpo:process}, and \ttl{evmpo:service} as a subclass of \ttl{evmpo:product}. Fig.~\ref{fig:emmo-evmpo} shows how, in terms of the \ttl{rdfs:subClassOf} relation, this class hierarchy can be integrated with the most closely related entities from the EMMO. It should be noted that the EMMO, and consequently also the integration as outlined in Fig.~\ref{fig:emmo-evmpo}, is work in progress and may therefore be subject to significant extensions and modifications in the future.

Consistency with the EVMPO, and by implication consistency with the EMMO (subsequent to a release of the first stable version of the EMMO~\cite{EMMC19}), is a requirement for all components and infrastructures that aim at interoperating within the European Virtual Marketplace Framework. This design facilitates that VIMMP, MarketPlace, and others can agree on the definition of paradigmatic entities as well as annotations which are either directly or indirectly related to paradigmatic entities, while any platform retains the option to extend its own semantic base as required. To remain interoperable within the European Virtual Marketplace Framework, any such additional works need to be connected semantically to the EVMPO. In this way, a multi-tier system of ontologies is established as follows, from the top (most general) to the lowest (most specific) level:
\begin{itemize}
   \item{} Top-level ontology: The top level is specified within the EMMO \cite{EMMC19}.
   \item{} Marketplace-level fundamentals: The fundamental categories and their most significant subclasses are specified within the EVMPO.
   \item{} Marketplace-level ontologies: A more detailed development of the EVMPO categories, with a focus on service and model interoperability, is covered by domain ontologies at the marketplace level \cite{HCSTSLABMGKSFBSC20}. The structure of this system of ontologies mostly follows the subdivision of the semantic space into fundamental paradigmatic categories as specified by the EVMPO, cf.~Section~\ref{sec:marketplace-level}.
   \item{} Subdomain-specific ontologies: These ontologies deal with subcategories; \eg, subdividing the semantic space covered by the VIMMP Software Ontology (VISO), specific modelling and simulation granularity levels are addressed by VISO-EL (electronic), VISO-AM (atomistic and mesoscopic), and VISO-CO (continuum) \cite{HNBCSCELNSSTVC20}.
\end{itemize}
The design of the marketplace-level ontologies is guided by two
principles: First, they should include classes and
relations that are suitable to describe
services, agents, and other aspects that are
relevant to infrastructures which are expected to interoperate within
the European Virtual Marketplace Framework \cite{HCSTSLABMGKSFBSC20}. Beside materials modelling
marketplaces, this prospectively includes data marketplaces, innovation
platforms, translation environments and simulation platforms as well as
any resources connected to such infrastructures (e.g., external
databases) and services associated with them.
As a second principle, the structure and scope of the individual marketplace-level
ontologies should follow the subdivision of the semantic space into fundamental
paradigmatic categories from the EVMPO; by implication, any entities
that cannot be subsumed under any of these categories need to be
classified as subclasses of \ttl{evmpo:annotation}.

The alignment of the eight marketplace-level ontologies
MACRO, MMTO, OSMO, OTRAS, VICO, VISO, VIVO, and VOV~\cite{HCSTSLABMGKSFBSC20}
with the fundamental paradigmatic categories from the EVMPO
is specified in Section \ref{subsec:paradigmatic}.
Section \ref{subsec:osmo-vov} summarizes
how models and the associated variables can be described in this context;
it introduces the example scenario that will be used in Section~\ref{sec:example}
to illustrate a possible way of matching terms from the marketplace-level ontologies
with terms from the EMMO.

\subsection{EVMPO fundamental paradigmatic categories}
\label{subsec:paradigmatic}

Below we list the fundamental paradigmatic categories from the EVMPO and their definitions,
including a description of selected subclasses of the fundamental
categories and the related marketplace-level ontologies~\cite{HCSTSLABMGKSFBSC20} from the VIMMP project:
\begin{enumerate}
   \item{} \ttl{evmpo:assessment}, \ie, a proposition on the accuracy or performance of an entity or an a expression of trust in an entity \\
      $\longrightarrow$ \quad \ttl{evmpo:endorsement\_assessment}, \ie, an assessment by which an entity is claimed to be good or to be fit for a certain purpose \\
      $\longrightarrow$ \quad \ttl{evmpo:requirement\_assessment}, \ie, an assessment concerning computational requirements (\eg, computing time, memory, or hardware and software prerequisites) \\
      $\longrightarrow$ \quad \ttl{evmpo:validity\_assessment}, \ie, an assessment concerning the uncertainty or error associated with a data item or with a model or simulation workflow by which data are generated \\
      \textit{Marketplace-level ontology:}~VIMMP Validation Ontology (VIVO)
   \item{} \ttl{evmpo:calendar\_event}, \ie, a meeting or activity which is scheduled or can be scheduled; this is defined to be equivalent with the entity \ttl{Vevent} from the W3C iCalendar ontology with time zones as datatypes (ICALTZD) \cite{CM05} \\
      \textit{Marketplace-level ontology:}~Ontology for Training Services (OTRAS)
   \item{} \ttl{evmpo:communication}, \ie, any message that is communicated \\
      $\longrightarrow$ \quad \ttl{evmpo:declaration}, \ie, a communication without a recipient \\
      $\longrightarrow$ \quad \ttl{evmpo:interlocution}, \ie, a communication with recipient(s) \\
      $\longrightarrow$ \quad \ttl{evmpo:statement}, \ie, an elementary communication that cannot be decomposed into multiple statements \\
      \textit{Marketplace-level ontology:}~VIMMP Communication Ontology (VICO)
   \item{} \ttl{evmpo:information\_content\_entity}, \eg, a journal article or a graph; this is defined to be equivalent with \ttl{IAO\_0000030}, labelled \textit{information content entity}, from the Information Artifact Ontology (IAO) \cite{Ceusters12} \\
      \textit{Marketplace-level ontology:}~Ontology for Training Services (OTRAS)
   \item{} \ttl{evmpo:infrastructure}, \ie, infrastructure of an EVMPO interoperable platform (\eg, related to data, hardware, and software) \\
      \textit{Marketplace-level ontologies:}~Marketplace-Accessible Computational Resource Ontology (MACRO), VIMMP Software Ontology (VISO)
   \item{} \ttl{evmpo:interpreter}, corresponding to the concept from Peirce's semiotics, which is based on the triad \textit{sign, interpreter, object}; this is defined to be equivalent with \ttl{emmo-semiotics:interpreter} \\
      $\longrightarrow$ \quad \ttl{evmpo:agent}, \ie, an interpreter that can interact with an infrastructure (\eg, with a virtual marketplace platform) \\
      \textit{Marketplace-level ontology:}~VIMMP Communication Ontology (VICO)
   \item{} \ttl{evmpo:material}, \ie, an amount of a physical substance (or mixture of substances) that is part of a more comprehensive real-world object~\cite{EMMC19}; this is defined to be equivalent with \ttl{emmo-material:ma\-terial}
   \item{} \ttl{evmpo:model}, \ie, an entity that represents a physical object or process by direct similitude or by capturing the relations between its properties in a mathematical framework; this is defined to be equivalent with \ttl{emmo-models:model} \\
      \textit{Marketplace-level ontologies:}~Materials Modelling Translation Ontology (MMTO), Ontology for Simulation, Modelling, and Optimization (OSMO), VIMMP Software Ontology (VISO)
   \item{} \ttl{evmpo:process}, \ie, the temporal evolution of one or multiple entities \\
      $\longrightarrow$ \quad \ttl{evmpo:business\_process}, \ie, an abstract procedural representation of economic relations \\
      $\longrightarrow$ \quad \ttl{evmpo:physical\_process}; this is defined to be equivalent with \ttl{emmo-processual:process} \\
      \textit{Marketplace-level ontology:}~Ontology for Simulation, Modelling, and Optimization (OSMO)
   \item{} \ttl{evmpo:product}, \ie, a good or service (which can be offered either on a virtual marketplace or off-site) \\
      \textit{Marketplace-level ontologies:}~Materials Modelling Translation Ontology (MMTO), Ontology for Training Services (OTRAS)
   \item{} \ttl{evmpo:property}, \ie, an entity that is determined by an observation process, involving a specific observer that perceives or measures it; this is defined to be equivalent with \ttl{emmo-properties:property} \\
      \textit{Marketplace-level ontology:}~VIMMP Ontology of Variables (VOV)
\end{enumerate}
The fundamental paradigmatic categories need not
be disjoint; \eg, \ttl{evmpo:ma}\-\ttl{terial} and \ttl{evmpo:product} overlap,
since a material can be manufactured with the intent
of selling it as a commodity, by which it becomes a product.

Concerning category 9, \nb, due to including business processes,
which are not necessarily processes in the sense of the EMMO, \ttl{evmpo:process}
is a proper superclass of \ttl{emmo-process:process}. To facilitate ontology alignment,
the subclass \ttl{evmpo:physical\_process} was introduced,
which is equivalent to \ttl{emmo-process:process}. Subclasses
of \ttl{evmpo:physical\_process} include \ttl{evmpo:manufacturing\_process},
\ie, a physical process that serves the production of a good,
and \ttl{evmpo:simulation}, \ie, a simulation workflow.

\subsection{Variables and model objects}
\label{subsec:osmo-vov}


The purpose of the VIMMP Ontology of Variables (VOV) consists in
organizing the variables (in a broad sense, including constants) that
appear in modelling and simulation, and to connect them to models and algorithms in which they
are involved as well as to model-related entities (\eg, interaction sites) to
which they are attached. In VISO, these model-related entities are referred to
as model objects (class \ttl{viso:model\_object}).

In particular, the marketplace-level ontologies VOV, VISO, and OSMO can be combined
to characterize models, algorithms, and workflows.
Two relations from VOV that serve this purpose are \ttl{vov:involves}
and \ttl{vov:has\_at}\-\ttl{tached}. A triple of the type \ttl{:X vov:involves :Y} describes
that there is a mathematical expression, extension, or algorithmic
formulation of \ttl{:X} that contains the variable or the model object \ttl{:Y};
therein, \ttl{:X} can be a model, a constituent part of a model (\eg,
a physical equation or materials relation), a solver (numerical implementation of a model),
or a physical entity that is modelled. A triple of the type \ttl{:X vov:has\_attached :Y} is
employed to associate a model object \ttl{:X} with a variable \ttl{:Y}.
In this way, VOV, VISO, and OSMO can be used to state that
a rigid two-site Mie potential is employed as model a for ammonia:

{\small\color{MidnightBlue}
\begin{verbatim}
molmod:AMMONIA a osmo:einecs_listed_material;
   osmo:has_ec_number "231-635-3"^^xs:string.  # identifies ammonia

molmod:NH3_POTENTIAL a osmo:materials_relation;
   osmo:has_aspect_paradigmatic_content molmod:AMMONIA;
   vov:involves molmod:NH3_RIGID_UNIT.

molmod:NH3_RIGID_UNIT a viso-am:rigid_object;
   viso:has_part molmod:NH3_SITE_A, molmod:NH3_SITE_B,
      molmod:NH3_SITE_COM.

molmod:NH3_SITE_A a viso-am:mie_site, viso-am:mass_site.  # Mie site 1
molmod:NH3_SITE_B a viso-am:mie_site, viso-am:mass_site.  # Mie site 2
molmod:NH3_SITE_COM a viso-am:structureless_object.  # centre of mass
\end{verbatim}}

The scenario given by the statements above will be employed below as an
example for illustrating the procedure of matching marketplace-level
ontologies with the EMMO. The problem in general and the present procedure
are introduced in Section~\ref{sec:combining}. In
Section~\ref{subsec:transposition}, the present
scenario is reexpressed in terms of classes and relations from the EMMO,
which inevitably results in a loss of information; in
Section~\ref{subsec:correspondences}, conceivable (``candidate'')
correspondences resulting from this transposition are evaluated and
reformulated, contributing to an ontology alignment.

Further triples can provide more information, \eg,
on the size parameter of the Mie interaction sites, which is
here specified to be $\sigma = 2.5764$ \AA:

{\small\color{MidnightBlue}
\begin{verbatim}
molmod:NH3_POTENTIAL
   osmo:has_aspect_object_content molmod:NH3_POTENTIAL_SITE_A.

molmod:NH3_POTENTIAL_SITE_A a osmo:materials_relation;
   osmo:has_aspect_object_content molmod:NH3_CONDITION_POT_A;
   vov:involves molmod:NH3_SITE_A.

molmod:NH3_CONDITION_POT_A a osmo:quantitative_condition;
   osmo:contains_predetermined_variable molmod:NH3_PARAMETER_SIG.

molmod:NH3_PARAMETER_SIG a osmo:unique_elementary;
   osmo:has_variable_name "sigma"^^xs:string;
   osmo:has_initial_elementary_value molmod:NH3_ELEMENTARY_SIG;
   osmo:has_variable_unit qudt-unit:Angstrom.

molmod:NH3_ELEMENTARY_SIG a osmo:elementary_value;
   osmo:is_decimal 2.5764.
\end{verbatim}}

Therein, the entity representing the unit \AA{} is obtained
from the system of ontologies for quantities, units, and datatypes (QUDT) \cite{ZLZP17}.

\section{Combining top-level and domain ontologies}
\label{sec:combining}

\subsection{Semantic heterogeneity}
\label{subsec:heterogeneity}

A major design goal for a top-level ontology (see Mascardi~\etal~\cite{MCR07}
for an illustrating comparison of popular ones) consists in
achieving the desired level of expressivity with a minimal repertoire
of basic terms and relations.
Obversely, to ensure interoperability at the marketplace level,
the employed ontologies need to capture detailed
characteristics of many particular services, models,
and interactions. Accordingly, the structure of the corresponding
semantic space at the lower level is comparably complex;
\eg, the marketplace-level ontologies from VIMMP
contain over 400 object class definitions (and, for technical reasons, over
800 additional definitions of property classes), over 250 definitions
of object properties (\ie, instances of \ttl{owl:ObjectProperty}),
and over 100 definitions of data properties (\ie, instances of \ttl{owl:DatatypeProperty}).
Therefore, by design, the EMMO needs to have a structure that is
substantially different from that of the marketplace-level ontologies \cite{HCSTSLABMGKSFBSC20}.
To ensure that the EMMO is consistently employed at all levels,
so that it can contribute to platform and service interoperability
as far as possible, the marketplace-level ontologies need to be
aligned with the EMMO.
Before returning to this specific problem, the present section summarizes
some of the related theoretical concepts.

In principle, semantic assets are designed to allow data integration and
overcome the data heterogeneity problem; in reality, 
semantic heterogeneity does arise, and it grows over time as resources
are added to the semantic web. This is known as the Tower of Babel problem \cite{HH06, Iliadis19}.
While some authors regard any presence of semantic heterogeneity as a
failure of semantic interoperability and hope for universal
agreements, others think that it is unavoidable and look
for strategies to deal with it. This may involve a standardized way
of documenting semantic assets, basic agreements on the approach to
ontology design (\eg, formal monism), and meta-ontological
formalizations of roles, procedures, and good practices (or best practices),
aiming at pragmatic interoperability \cite{AS10, HCSTSKK20, SD20, SMD06}.
For this approach, the challenge consists in agreeing and
specifying how the semantic space is structured, documented, and employed
in practice; by raising the domain for which universal agreements are
pursued from the ontological level to the meta-ontological level, ``the
Tower of Babel becomes a Meta-Tower of Babel'' \cite{Varzi19}.

As a consequence, semantic heterogeneity is seen as a necessary property
of the semantic web, and ontology matching and integration become basic
features of its successful mode of operation, rather than an expression
of incompleteness. Options for implementing such a mode of operation have been
extensively discussed in the literature, first for schemas and then for
ontologies; for example, \cf~Noy~\cite{Noy04}, Euzenat and Shvaiko~\cite{ES13},
the Ontology Matching website \cite{SE19}, and the material from
a series of events organized by the Ontology Alignment Evaluation
Initiative (OAEI) \cite{Jimenez19}.
The common challenge is how to make use of the knowledge represented in two
ontologies, which can differ at various levels (language used, expressivity, modelling
paradigm, \etc). Typically, such challenges arise if there is an overlap in the
domains of knowledge addressed by multiple ontologies, such that data annotated
in diverse ways need to be combined and processed together, or if
a platform employs multiple domain ontologies that are based on different top-level ontologies.
Typical applications include, \eg, simultaneous
querying of multiple knowledge bases or, as addressed here, the
transposition of semantic content from a source ontology $s$ to a target ontology $t$.

\subsection{Ontology alignment}
\label{subsec:alignment}

Requirements generally differ between tasks occurring at runtime
(emphasis on efficiency) and at design time (emphasis on completeness).
Similarly, the desired outcome is not always the same: As opposed to ontology
integration, \ie, the creation of a merged (integrated) ontology,
\textit{ontology matching} is a process that yields a set of correspondences:
An ontology alignment \cite{ES13, SE13}. This is known as the ontology matching problem (OMP), for which a
great variety of strategies and algorithms have been devised \cite{AR14, ES13, MA12, MMCMGP17, SE13}.
The correspondences can be formulated, with increasing expressive power, as follows:
\begin{enumerate}
   \item{} As tuples $(\sigma, \tau, \omega)$, where $\sigma$ and $\tau$ are classes or relations
      from the source and target ontologies $s$ and $t$, respectively,
      and $\omega \in \{\sqsubseteq, \sqsupseteq, \equiv\}$ is a logical operator
      indicating whether $\sigma$ is a subclass or subproperty (\ie, $\sigma \sqsubseteq \tau$)
      or a superclass or superproperty of $\tau$ (\ie, $\sigma \sqsupseteq \tau$), or whether
      the two are equivalent (\ie, $\sigma \equiv \tau$). If a statistical approach is
      employed, the tuple may contain an additional
      metadata item indicating the probability that the operator can be
      applied; \cf~Shvaiko and Euzenat~\cite{SE13} for a discussion of such approaches.

   \item{} As OWL statements (\eg, in TTL format) or equivalently as formul\ae{} in OWL description logic (OWL DL);
      where viable, this approach is preferable in the present context, since it can be implemented
      immediately in any technical infrastructure where the marketplace-level
      ontologies are used.

   \item{} In more expressive logics, \eg, first-order logic or
      description logics that extend OWL DL \cite{BHLS17},
      or by rewriting rules constituting, \eg, a graph transformation system~\cite{KNPR18}. While these formalisms
      are undecidable in general, their particular application to the OMP might
      restrict formul\ae{} and rules in a way that keeps them decidable and
      tractable computationally; this will be addressed by future work, since as
      discussed below, OWL DL is insufficient to capture a number of typical and relevant cases.
\end{enumerate}
For the present purpose, we will assume that the source ontology, which here
is a marketplace-level ontology, is more expressive semantically, and less abstract;
the EMMO will be the target ontology.

Since ontologies can easily involve hundreds of entities, automatic and semi-automatic tools with different levels of user involvement
have been created to perform ontology matching;
see, \eg, Tab.~2 from Mart\'inez and Aldana \cite{MA12} for a list of algorithms, and Tab.~1 from Shvaiko and Euzenat \cite{SE13} for a list of tools.
The strategy they use to find correspondences is to look for similarities between
entities (classes, relations, individuals), considering various aspects: Label and documentation, using exact or approximate text matching (terminological/linguistic aspect); subclasses and superclasses,
relations, their domain and ranges (structural aspect); instances of classes
(extensional aspect);
data type (constraint-based aspect); class descriptions in the sense of rules
(semantic aspect) \cite{Noy04, SE13}. 
Different matchers can be combined, with weights and thresholds, to build a
meta-matching approach~\cite{MA12};
an API to describe alignments has been proposed as well~\cite{Euzenat04}.

To be able to compare possible candidate correspondences, distances and similarities measures can be defined~\cite{MA12}.
The quality of the overall alignment can also be evaluated, using precision
(a measure of correctness)
and recall (a measure of completeness); a combination of the two
is called F-measure~\cite{SE13}. These concepts were borrowed from the
area of data retrieval; in the context of ontology alignment,
they can be defined as follows:
``given a reference alignment $R$, the precision of an alignment $A$ is
$P(A,R) = |R \cap A| / |A|$, and the recall
is given by $R(A,R) = |R \cap A| / |R|$'' (Euzenat~\cite{Euzenat07}, Definition 9). Note that for this
purpose of evaluation and comparison of alignments, a reference alignment, for
example compiled by a human, has
to be provided.
To support the matching process, external resources can be used, \eg,
a higher-level ontology operating at a moderate level of abstraction~\cite{MLR10}
or a linguistic resource such as WordNet~\cite{Miller95}.




Applying the OMP to the particular problem at hand, this work aims
at establishing a limited semantic space which is aligned between
the two levels, serving both as a proof of concept and as a
basis for future work that will further extend the aligned space.
In the present case, the concerned ontologies are all in an early
stage of development. As a consequence, substantial corpora of triples
formulated in any of the ontologies do not exist; any attempt
at matching the marketplace-level ontologies with the EMMO needs to be
based on limited sets of triples that were either created for this
special purpose or to guide the marketplace platform design.
This precludes any automated approach to ontology matching. Instead,
the present strategy is purely based on (human-)asserted
correspondences of classes and relations.
It can be summarized as follows; we assume that in the beginning,
a description in terms of the source ontology $s$ is available
for a scenario that is well understood by the ontologist who 
carries out the matching:
\begin{enumerate}
   \item{}
      Reexpress the scenario in terms of the target ontology $t$;
      observe how source-ontology terms $\sigma^\circ_i$, \ie,
      classes, relations, or more complex expressions, are mapped
      to target-ontology terms $\tau^\circ_i$, where $i \in \mathbb{N}$ is an index.
      These instances, $\sigma^\circ_i \mapsto \tau^\circ_i$, yield
      candidate correspondences $\sigma^\circ_i \sqsubseteq^? \tau^\circ_i$,
      each of which is further addressed by the subsequent steps.

   \smallskip

   \item{}
      If the candidate correspondence is valid, $\sigma^\circ_i \sqsubseteq \tau^\circ_i$,
      proceed to the next step without any change, \ie, $\sigma_i' = \sigma^\circ_i$,
      $\tau_i' = \tau^\circ_i$. If it is invalid, $\sigma^\circ_i \not\sqsubseteq \tau^\circ_i$,
      relax it such that $\sigma_i' \sqsubseteq \tau_i'$ becomes valid by
      \begin{itemize}
         \item{} replacing $\tau^\circ_i$ with a more general
            term $\tau_i' \sqsupseteq \tau^\circ_i$ ($\tau$-generalization), or
         \item{} replacing $\sigma^\circ_i$ with a more specific
            term $\sigma_i' \sqsubseteq \sigma^\circ_i$ ($\sigma$-refinement).
      \end{itemize}
      If this attempt reduces the correspondence to a trivial
      statement, discard the candidate correspondence.

   \smallskip

   \item{}
      Strengthen the correspondence as far as possible by
      \begin{itemize}
         \item{} replacing $\sigma_i'$ with a more general term
            $\sigma_i'' \sqsupseteq \sigma_i'$ ($\sigma$-generalization),
         \item{} replacing $\tau_i'$ with a more specific term
            $\tau_i'' \sqsubseteq \tau_i'$ ($\tau$-refinement), or
         \item{} replacing the subclass/subproperty operator $\sqsubseteq$
            with the equivalence operator $\equiv$ ($\sigma$-$\tau$-identification).
      \end{itemize}
      The outcome has the form $\sigma_i'' \sqsubseteq \tau_i''$, or
       $\sigma_i'' \equiv \tau_i''$ if the operator has been replaced.

   \smallskip

   \item{}
      Express the correspondence in OWL (\eg, TTL notation) or OWL DL
      if possible. If it cannot be expressed in OWL, attempt to simplify
      it, \eg, by relaxing it again as far as necessary; if that is
      impossible without reducing the correspondence to a trivial statement,
      the candidate is discarded.
\end{enumerate}
Step 4 can easily be adjusted in cases where there is no need to
introduce the alignment immediately into an OWL based architecture.

This procedure is not meant to be automated; in particular, for the present problem
of matching domain ontologies with the EMMO pre-release version, it requires a close reading of
definitions from the source and target ontologies and the available
documentation to assess how the statements must be relaxed and
strengthened, respectively, to be valid and as powerful as possible.



\section{Example: Intermolecular pair potential}
\label{sec:example}

\subsection{Reexpressing the scenario in the target ontology}
\label{subsec:transposition}

The following remarks refer to the EMMO pre-release development version 0.9.10~\cite{EMMC19};
the precise way of mapping individuals and relations between
them from the marketplace-level ontology representation to an EMMO representation
will be subject to change as all of the involved ontologies are further developed.
However, the main purpose of the present example consists not in its result, but
in taking the first steps toward ontology alignment within the European Virtual
Marketplace Framework, evaluating to what extent the method
proposed in Section~\ref{subsec:alignment} is suitable for this purpose,
and providing an orientation for future work on this basis.

Mapping the first part of the example from Section~\ref{subsec:osmo-vov} to
entities from EMMO version 0.9.10 yields:

{\small\color{MidnightBlue}
\begin{verbatim}
molmod:AMMONIA a emmo-material:material.

molmod:NH3_POTENTIAL a emmo-models:material_relation.
molmod:AMMONIA emmo-models:has_model [
      a emmo-models:physics_based_model;  
      emmo-mereotopology:has_proper_part molmod:NH3_POTENTIAL,
         molmod:NH3_RIGID_UNIT
   ].

molmod:NH3_RIGID_UNIT a emmo-graphical:symbolic, emmo-semiotics:sign;
   emmo-mereotopology:has_proper_part molmod:NH3_SITE_A,
      molmod:NH3_SITE_B, molmod:NH3_SITE_COM.

molmod:NH3_SITE_A a emmo-graphical:symbol.
molmod:NH3_SITE_B a emmo-graphical:symbol.
molmod:NH3_SITE_COM a emmo-graphical:symbol.
\end{verbatim}}

To keep the example above and the present discussion readable,
EMMO entities are represented by their labels rather than their
identifiers; \eg, we write \ttl{emmo-models:has\_model} to represent
the relation that has the identifier \ttl{emmo-models:EMMO\_24c71baf\_6db6\_48b9\_86c8\_8c70cf36db0c}
and the label \ttl{"has\_model"}.

The transposed version above does not include any entity corresponding to
the elementary data item from the source-ontology
version of the example, \ie, the string \ttl{"231-635-3"}, since
the present target ontology does not include any data properties and elementary
data items; in the EMMO, strings,
numbers, \etc, are \ttl{emmo-graphical:symbolic} individuals rather than elementary
data items, and all relations that are defined in the EMMO are object properties.
Accordingly, it would be possible to introduce an EMMO
individual representing \ttl{"231-635-3"}; however,
since it is impossible in OWL for an elementary data item to be
the \textit{same as} (\ttl{owl:sameAs}) an
individual, this would not yield any formal correspondence at
the ontological level. The same, consequently, applies
to the datatype properties from the marketplace-level ontologies:
They cannot be matched with any of the relations from the EMMO;
the EMMO only contains object properties, which cannot be \textit{equivalent}
(\ttl{owl:equivalentProperty}) to datatype properties.

Considering the present example, it should
be noted that entities such as the pair potential
or the molecular centre of mass from the example are not actual features
of fluid ammonia, but of a molecular model for ammonia.
For the transposition of the scenario, we assume that in an appropriate
formal language for molecular models, interaction sites and other
\textit{structureless objects} (\ttl{viso-am:structureless\_object}) are syntactically irreducible
elementary symbols and hence \ttl{emmo-graphical:symbol} individuals by the EMMO definition~\cite{EMMC19},
whereas a rigid unit that contains multiple sites is an entity
composed of multiple symbols (in the theory of formal languages, a \textit{word})
and hence a \ttl{emmo-graphical:symbolic} entity, which in the EMMO is
defined as a composition of symbols, including a single symbol as a special case~\cite{EMMC19}.
The specification of such a formal language, which would be unproblematic,
is beyond the scope of the present work. We merely note that by the
EMMO definition, symbols ``of a formal language'' are the smallest
irreducible parts of symbolic entities, which depends on the way in which
the formal language is specified; it is certainly possible to refer
to interaction sites and other structureless objects
by elementary symbols ($a$, $b$, $c$, \dots, or special symbols or
elementary keywords devised for this purpose).

By its foundation on Peircean semiotics \cite{Peirce91}, the EMMO applies a clear
distinction between physical objects and the signs for them;
thereby, symbolic entities (including irreducible symbols) need not
be signs; they only become signs
if they represent existing physical objects in the eyes of an
interpreter. (Here, the interpreter, who is
not explicitly represented in the example, might be a
provider or user of the molecular model.)
Moreover, due to the interpretation of mereotopology~\cite{Varzi96}
underlying the EMMO, any existing (``real world'') physical objects
must be four-dimensional spacetime entities; \ie, they must correspond
to trajectories of contiguous volumes that can be tracked over time.
Accordingly, the definition of the top entity from EMMO
version 0.9.10, \ttl{emmo-mereotopology:emmo}, states: ``A
real world object is then a 4D topological sub-region of the
universe. [\dots] It follows that, for the EMMO, real world objects
of dimensionality lower than 4D do not exist (\eg{} surfaces, lines)'' \cite{EMMC19}.
If a symbolic entity is used to refer to any lower-dimensional structure,
it is not a \ttl{emmo-semiotics:sign} (nor an \ttl{emmo-models:model}),
since it does not represent a ``real-world object'' in the sense
attributed to this term by the EMMO.

Hence, in particular, \ttl{molmod:NH3\_RIGID\_UNIT}, the rigid unit that contains
the interaction sites of the molecular model, is a sign for an ammonia molecule.
It is debatable whether it is also a model in the sense of the EMMO, which
requires ``direct similitude'' with the represented object or a mathematical
formalization of its behaviour \cite{EMMC19}; here, we do not classify it as a model.
The two Mie sites \ttl{molmod:NH3\_SITE\_A} and \ttl{molmod:NH3\_SITE\_B}, however,
are only symbols (and part of a sign), since they do not represent any particular
part of the molecule or any other part of spacetime specifically.
The pair potential \ttl{molmod:NH3\_POTENTIAL} might be
understood to represent the potential energy from the pairwise
interaction between two NH$_3$ molecules; however, this does
not constitute semiosis in the sense of the EMMO, since the
referenced object is not a 4D entity, but a hypothetical additive contribution
to a scalar quantity. Consequently, \ttl{molmod:NH3\_POTENTIAL}
is \textit{not} an \ttl{emmo-models:model} individual; however, it
is a part of such an individual, since the combination of all
pair potentials involved in modelling a system yields a parameterization
of a classical-mechanical equation of motion that describes the trajectory
of the system, which is a spacetime individual in the sense of the EMMO.
In the same way, \ttl{molmod:NH3\_SITE\_COM}, which would represent the centre
of mass of the molecule, cannot be a sign in the sense of the EMMO:
The centre of mass is a point (not a volume), and its trajectory
over time is one-dimensional rather than four-dimensional.
However, \ttl{molmod:NH3\_SITE\_COM} is part of a sign, by belonging to \ttl{molmod:NH3\_RIGID\_UNIT}.

\subsection{Construction of ontological correspondences}
\label{subsec:correspondences}

The procedure from Section~\ref{subsec:alignment} will now be followed
to create the fragment of an ontology alignment between marketplace-level ontologies and the
EMMO. By reexpressing the source-ontology (OSMO, VISO, and VOV)
scenario from Section~\ref{subsec:osmo-vov} in the target ontology (EMMO) as in Section~\ref{subsec:transposition},
classes and relations are mapped to each other (step 1)
\begin{equation}
   \sigma_i^\circ ~ \mapsto ~ \tau_i^\circ  \quad  \textnormal{for} ~ 1 \leq i \leq 10,
\label{eqn:step1}
\end{equation}
where the initial source terms are
\begin{eqnarray}
   \sigma_1^\circ ~ & ~ = ~ & ~ \ttl{osmo:einecs\_listed\_material}, \label{eqn:step1source} \\
   \sigma_2^\circ ~ & ~ = ~ & ~ \ttl{osmo:materials\_relation}, \nonumber \\
   \sigma_3^\circ ~ = ~ \sigma_4^\circ ~ & ~ = ~ & ~ \ttl{viso-am:rigid\_object}, \nonumber \\
   \sigma_5^\circ ~ & ~ = ~ & ~ \ttl{viso-am:mie\_site}, \nonumber \\
   \sigma_6^\circ ~ & ~ = ~ & ~ \ttl{viso-am:mass\_site}, \nonumber \\
   \sigma_7^\circ ~ & ~ = ~ & ~ \ttl{viso-am:structureless\_object}, \nonumber \\
   \sigma_8^\circ ~ & ~ = ~ & ~ \ttl{viso:has\_part}, \nonumber \\
   \sigma_9^\circ ~ & ~ = ~ & ~ \ttl{vov:involves}, \nonumber \\
   \sigma_{10}^\circ ~ & ~ = ~ & ~ \ttl{osmo:has\_aspect\_paradigmatic\_content}, \nonumber 
\end{eqnarray}
and the initial target terms are
\begin{eqnarray}
   \tau_1^\circ ~ & ~ = ~ & ~ \ttl{emmo-material:material}, \label{eqn:step1target} \\
   \tau_2^\circ ~ & ~ = ~ & ~ \ttl{emmo-models:material\_relation}, \nonumber \\
   \tau_3^\circ ~ & ~ = ~ & ~ \ttl{emmo-semiotics:sign}, \nonumber \\
   \tau_4^\circ ~ & ~ = ~ & ~ \ttl{emmo-graphical:symbolic}, \nonumber \\
   \tau_5^\circ = \tau_6^\circ = \tau_7^\circ ~ & ~ = ~ & ~ \ttl{emmo-graphical:symbol}, \nonumber \\
   \tau_8^\circ ~ & ~ = ~ & ~ \ttl{emmo-mereotopology:has\_proper\_part}, \nonumber \\
   \tau_9^\circ ~ & ~ = ~ & ~ (\ttl{emmo-mereotopology:has\_proper\_part})^- \nonumber \\
      & & \quad \quad \circ ~ \ttl{emmo-mereotopology:has\_proper\_part}, \nonumber \\
   \tau_{10}^\circ ~ & ~ = ~ & ~ (\ttl{emmo-models:has\_model} ~ \circ \nonumber \\
      & & \quad \quad \ttl{emmo-mereotopology:has\_proper\_part})^-. \nonumber
\end{eqnarray}
Therein, $p^-$ denotes the inverse relation to $p$, and $p \circ q$ denotes
the chain product of the relations $p$ and $q$.

By evaluating the respective candidate correspondences (step 2) on the basis of
the definitions from the marketplace-level ontologies and the EMMO,
$\sigma_i^\circ \sqsubseteq \tau_i^\circ$ is found to be valid for
$i \in \{1, 2, 4, 5, 6, 7\}$; hence, $\sigma'_i = \sigma_i^\circ$
and $\tau'_i = \tau_i^\circ$ for these $i$.
For $i = 3$, no non-trivial and valid relaxation can be found:
A rigid object need not be a sign according to the EMMO, and whether it is one or not
depends on its understanding by an interpreter. For the associated semiotic
processes, a clear and simple rule cannot be formulated;
accordingly, this candidate is discarded. The correspondences between
relations can be relaxed by $\tau$-generalization to
\begin{equation}
   \tau'_8 ~ = ~ \ttl{emmo-mereotopology:has\_part},
\end{equation}
with $\sigma'_8 = \sigma_8^\circ$, 
and by $\sigma$-refinement to
\begin{eqnarray}
   \sigma_{9}' ~ & ~ = ~ & ~ \ttl{vov:involves} ~ \sqcap ~ [\ttl{osmo:materials\_relation}]^\bullet \\
      && \quad \quad \sqcap ~ ^\bullet[\ttl{viso:model\_object}], \nonumber \\
   \sigma_{10}' ~ & ~ = ~ & ~ \ttl{osmo:has\_aspect\_paradigmatic\_content} \nonumber \\
    & & \quad \quad \sqcap ~ [\ttl{osmo:materials\_relation}]^\bullet ~ \sqcap ~ ^\bullet[\ttl{evmpo:material}], \nonumber
\end{eqnarray}
with $\tau'_{9} = \tau_{9}^\circ$ and
$\tau'_{10} = \tau_{10}^\circ$, where $[c]^\bullet$ is the relation that
holds whenever the subject is an individual of class $c$ (irrespective of the object),
$^\bullet{}[c]$ is the relation that holds whenever the object is an individual
of class $c$ (irrespective of the subject), and $\sqcap$ is the
intersection operator, applied to relations.
In this way, $\sigma_i' \sqsubseteq \tau_i'$ is valid
for all the remaining cases (\ie, $i \neq 3$).

Strengthening the correspondences (step 3) yields
\begin{equation}
   \ttl{evmpo:material} ~ \equiv ~ \ttl{emmo-material:material}
\label{eqn:step3case1}
\end{equation}
by $\sigma$-generalization and $\sigma$-$\tau$-identification for $i = 1$,
\begin{equation}
   \ttl{osmo:materials\_relation} ~ \equiv ~ \ttl{emmo-models:material\_relation}
\label{eqn:step3case2}
\end{equation}
by $\sigma$-$\tau$-identification for $i = 2$,
\begin{equation}
   \ttl{viso:model\_object} ~ \sqsubseteq ~ \ttl{emmo-graphical:symbolic}
\label{eqn:step3case4}
\end{equation}
by $\sigma$-generalization for $i = 4$,
\begin{equation}
   \ttl{viso-am:structureless\_object} ~ \sqsubseteq ~ \ttl{emmo-graphical:symbol}
\label{eqn:step3case567}
\end{equation}
for $i \in \{5, 6\}$ by $\sigma$-generalization and for $i = 7$ unchanged,
\begin{equation}
   \ttl{viso:has\_part} ~ \sqsubseteq ~ \ttl{emmo-mereotopology:has\_part}, \label{eqn:step3case8} \\
\end{equation}
for $i = 8$ unchanged,
\begin{eqnarray}
   & & \Big(\ttl{vov:involves} \label{eqn:step3case9} \\
      & & \quad \quad \sqcap ~ [\ttl{osmo:materials\_relation}]^\bullet ~ \sqcap ~ ^\bullet[\ttl{viso:model\_object}] \Big) \nonumber \\
      & & \quad \quad \quad \quad \sqsubseteq \quad \Big(\Big([\ttl{emmo-models:model}]^\bullet \nonumber \\ 
      & & \quad \quad \quad \quad \quad \quad \sqcap ~ \ttl{emmo-mereotopology:has\_proper\_part}\Big)^- \nonumber \\
      & & \quad \quad \quad \quad \quad \quad \quad \quad \circ ~ \ttl{emmo-mereotopology:has\_proper\_part}\Big), \nonumber
\end{eqnarray}
for $i = 9$ by $\tau$-refinement, and
\begin{eqnarray}
   & & \Big(\ttl{osmo:has\_aspect\_paradigmatic\_content} \label{eqn:step3case10} \\
      & & \quad \quad \quad \sqcap ~ [\ttl{osmo:materials\_relation}]^\bullet ~ \sqcap ~ ^\bullet[\ttl{evmpo:material}]\Big) \nonumber \\
      & & \quad \quad \quad \quad \quad \quad \sqsubseteq \quad \Big(\ttl{emmo-models:has\_model} ~ \circ \nonumber \\
      & & \quad \quad \quad \quad \quad \quad \quad \quad \quad \ttl{emmo-mereotopology:has\_proper\_part}\Big)^-, \nonumber
\end{eqnarray}
for $i = 10$ unchanged.

Eqs.~(\ref{eqn:step3case1}) to (\ref{eqn:step3case8}) can be expressed in OWL (step 4):

{\small\color{MidnightBlue}
\begin{verbatim}
evmpo:material owl:equivalentClass emmo-material:material.
osmo:materials_relation owl:equivalentClass emmo-models:material_relation.
viso:model_object rdfs:subClassOf emmo-graphical:symbolic.
viso-am:structureless_object rdfs:subClassOf emmo-graphical:symbol.
viso:has_part rdfs:subPropertyOf emmo-mereotopology:has_part.
\end{verbatim}}

Therein, the first statement was already included in the EVMPO-EMMO integration,
\cf~Section~\ref{subsec:evmpo} and Fig.~\ref{fig:emmo-evmpo}, whereas the subsequent
statements contribute to aligning the marketplace-level ontologies with the EMMO.

The present example also illustrates the limitations due to using OWL
for the purpose of formalizing the alignment.
Eqs.~(\ref{eqn:step3case9}) and (\ref{eqn:step3case10}) exceed the expressive
capacity of OWL; however, the meaning of these correspondences is intuitively straightforward,
and similar cases occur often in practice: Eq.~(\ref{eqn:step3case9}),
from $i = 9$, states that ``if \ttl{:X vov:involves :Y}, where
\ttl{:X} is a \ttl{osmo:materials\_relation} and \ttl{:Y} is a \ttl{viso:model\_object},
then there is an \ttl{emmo-models:model} that relates to both of them
by \ttl{emmo-mereo}\-\ttl{topology:has\_proper\_part},'' \ie, they are both part of
the same model. Eq.~(\ref{eqn:step3case10}),
from $i = 10$, expresses the rule that ``if \ttl{:X osmo:has\_}\-\ttl{aspect\_para}\-\ttl{digmatic\_content :Y},
where \ttl{:X} is a \ttl{osmo:materials\_rela}\-\ttl{tion} and \ttl{:Y} is a \ttl{evmpo:material},
then there is an individual \ttl{:Z} such that \ttl{:Y em}\-\ttl{mo-models:has\_}\-\ttl{model :Z}
and \ttl{:Z emmo-mereotopology:has\_proper\_}\-\ttl{part :X}.''
Statements like these can easily be formulated in first-order logic,
extensions of OWL DL by additional operators (see above), or by
graph transformation systems~\cite{KNPR18} which would apply
to graph representations of the scenarios (see the Appendix).

\section{Conclusion}
\label{sec:conclusion}

Data technology for CME/ICME services and platforms needs to aim at
FAIR data management~\cite{Bicarregui16} and semantic interoperability with respect to concrete
software and data infrastructures, \eg, materials modelling marketplaces,
for which it is crucial to develop marketplace-level
domain ontologies by which the associated models, tools, infrastructures,
and workflows, can be documented in detail. The EMMO,
on the other hand, aims at capturing
materials modelling and characterization in general; beyond this, it
is expected to serve as an entirely domain-independent top-level ontology.
The present work shows that the gap between the top level and the
domain level can indeed be bridged, even where, as it is the case here,
the involved ontologies are at a relatively early stage of development.

An ontology alignment, once established, becomes community know\-ledge
and remains available for future use; we suggest to include these alignment
statements in a TTL file that should be distributed together with the
EVMPO and the marketplace-level ontologies~\cite{HCSTSLABMGKSFBSC20}.
As such knowledge is accumulated over time, the EMMO becomes more accessible
to its prospective community of users, not all of whom can be expected
to familiarize themselves with all intricacies of the philosophical
conceptualization underlying the EMMO classes and relations.
Domain ontologies, however, are comparably close to the language
employed in a particular field and the intuition of domain experts.
Therefore, ontology matching, even fragment by fragment, successively
contributes to creating the necessary link between domain-specific
expertise and the EMMO as an overarching interoperability standard.

To illustrate this, it is improbable that among developers of
molecular simulation software, a majority will at any point be aware of
the circumstance that a Lennard-Jones interaction site can be
documented as an \ttl{emmo-graphical:symbol} individual, or that a
multi-site rigid-unit part of a molecular model is best represented as an
\ttl{emmo-graphical:symbolic}. Instead, a CME/ICME domain expert might
find it more easy to recognize that \ttl{viso-am:lj\_site} and
\ttl{viso-am:rigid\_object}, respectively, are appropriate concepts.
Then, from VISO~\cite{HNBCSCELNSSTVC20} and the present ontology alignment,
\begin{eqnarray}
   \ttl{viso-am:lj\_site} ~ &\sqsubseteq& ~ \ttl{viso-am:structureless\_object} \\
      &\sqsubseteq& ~ \ttl{emmo-graphical:symbol}, \nonumber \\
   \ttl{viso-am:rigid\_object} ~ &\sqsubseteq& ~ \ttl{viso:model\_object} \\
      &\sqsubseteq& ~ \ttl{emmo-graphical:symbolic}, \nonumber
\end{eqnarray}
the EMMO representation of these entities can be deduced correctly.

To make this approach and the EMMO-based semantic interoperability architecture
as a whole more viable, future work should address mechanisms by which
alignment rules can be included if they are beyond the expressive capacity
of OWL DL; \eg, graph rewriting rules could be applied to representations
of scenarios as graphs, building on the work in progress documented in the Appendix.
Furthermore, a community-approved procedure needs to be
established for suggesting, evaluating, and disseminating
ontology alignments that connect the EMMO to domain knowledge.

\begin{acknowledgement}
The authors thank N.~Adamovic, W.~L.~Cavalcanti, \AA.~Ervik, A.~Hashibon, and E.~A.~M\"uller for fruitful discussions.
The co-authors Y.B., E.G., G.G., G.M., and G.J.S.\ acknowledge funding from the European Union's Horizon 2020 research and innovation programme under grant agreement no.\ 723867 (EMMC-CSA), the co-authors Y.B., G.G., G.M., and G.J.S.\ under grant agreement no.\ 760173 (MarketPlace), the co-authors S.C., G.G., M.T.H., G.M.\ under grant agreement no.\ 760907 (VIMMP), and the co-author E.G.\ under grant agreement no.\ 814492 (SimDOME).
\end{acknowledgement}

\section*{Appendix: Graph representation of EMMO scenarios}
\addcontentsline{toc}{section}{Appendix}

\begin{figure}[p]
\centering
\includegraphics[width=12cm]{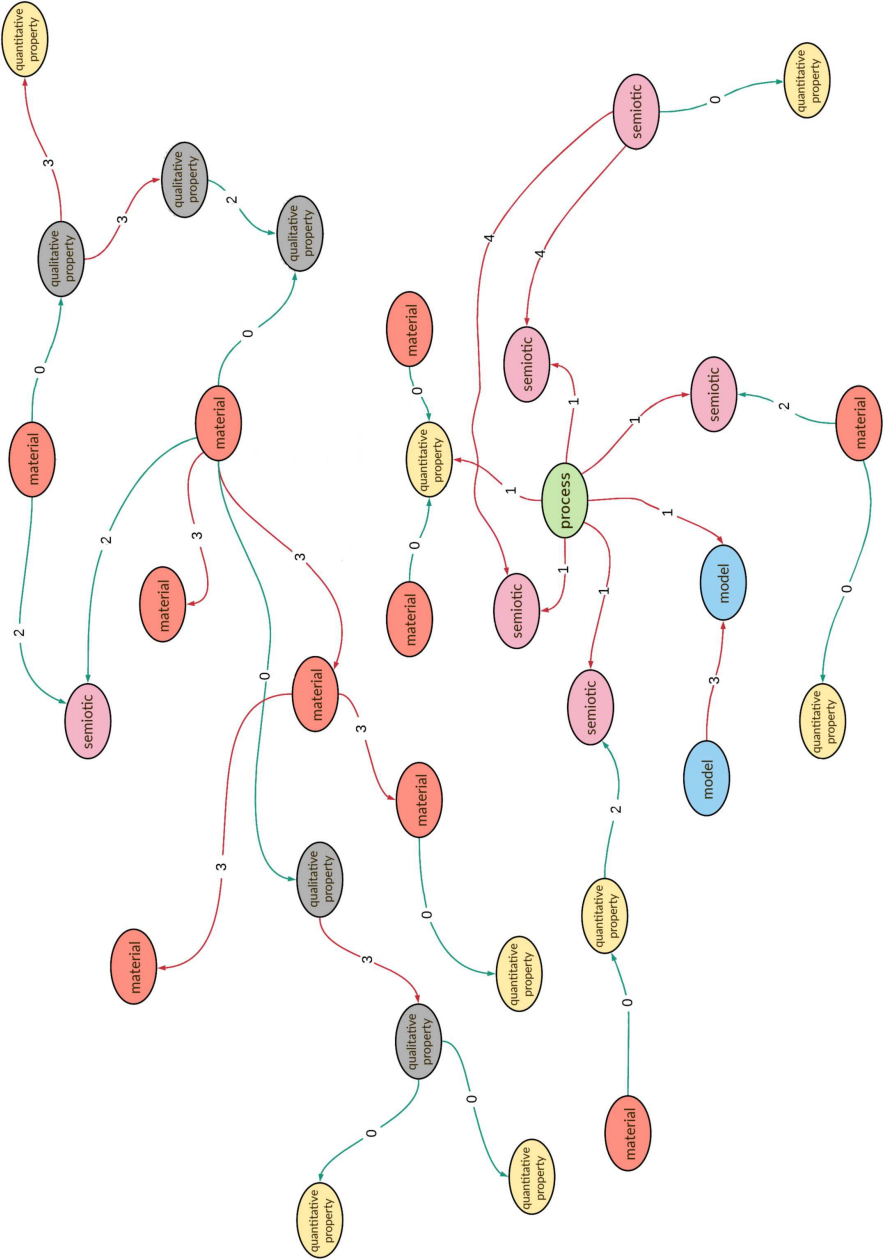}
\caption{Graph representation of EMMO individuals and relations from a scenario describing a simulation workflow for crystal structure prediction; \cf~Tab.~\ref{tab:edge-labels} for the edge labels. This diagram was created using Lucidchart~\cite{Lucidchart20}.}
\label{fig:lucidchart-model}
\end{figure}

Using the EMMO Python API~\cite{EMMC19} in combination with the Lucidchart online tool~\cite{Lucidchart20},
scenarios represented as a graph (\eg, see Fig.\ \ref{fig:lucidchart-model}) can be
converted automatically to a list of Python classes consistent with the Owlready2
format \cite{Lamy17} which facilitates further processing by the EMMO API.
The complete procedure involves performing the following steps:
\begin{enumerate}
   \item{} Assign numerical labels to concepts and relations (\eg, see Tab.~\ref{tab:edge-labels}).
   \item{} Export the flowchart data from Lucidchart~\cite{Lucidchart20} into CSV format.
   \item{} Insert parameters in the \texttt{parameters.py} module file~\cite{EMMC19}.
   \item{} Execute the \texttt{CSP\_ontology.py} script~\cite{EMMC19} to generate the list of classes for Owlready2~\cite{Lamy17}.
\end{enumerate}
For further details, we refer to the EMMO Lucidchart
documentation~\cite{Mogni19} and the upcoming release version of the EMMO.

\begin{table}[h]
\caption{Codes (edge labels) employed for the EMMO relations in the present graph representation.}
\label{tab:edge-labels}
   \begin{tabular}{cl}
   \hline
      code ~ & ~ EMMO relation name and description, \cf~Ghedini~\etal~\cite{EMMC19} \\
   \hline
        0  ~ & ~ \ttl{emmo-properties:has\_property} \\
           ~ & ~ a property is a ``sign that stands for an object that the 'interpreter' \\
           ~ & ~ perceived through a well defined observation process'' \cite{EMMC19}, \eg, an \\
           ~ & ~ experimental measurement process; \ttl{has\_property} relates the observed \\
           ~ & ~ object to the outcome of that process \\
        1  ~ & ~ \ttl{emmo-processual:has\_proper\_participant} \\
           ~ & ~ relates a process to an entity that participates in the process; irreflexive \\
           ~ & ~ relation (as opposed to \ttl{has\_participant}) \\
        2  ~ & ~ \ttl{emmo-semiotics:has\_sign} \\
           ~ & ~ relates an object to a sign that refers to it by participation in semiosis \\
        3  ~ & ~ \ttl{emmo-mereotopology:has\_proper\_part} \\
           ~ & ~ irreflexive parthood relation \\
        4  ~ & ~ \ttl{emmo-mereotopology:has\_spatial\_part} \\
           ~ & ~ ``relation that isolates a proper part that extends itself in time within the \\
           ~ & ~ lifetime of the whole, without covering the full spatial extension of the 4D \\
           ~ & ~ whole (\ie{} is not a temporal part)'' \cite{EMMC19} \\
   \hline
   \end{tabular}
\end{table}

\bibliography{emmo-integration-iciam}
\end{document}